# Inter-laboratory comparison of a WDS-EDS quantitative x-ray microanalysis of a metallic glass

Philippe Jonnard*[1,2], François Brisset[3], Florence Robaut[4], Guillaume Wille[5], Jacky Ruste[6]

[1] Sorbonne Universités, UPMC Univ Paris 06, Laboratoire de Chimie Physique-Matière et Rayonnement, 11 rue Pierre et Marie Curie, F-75231 Paris cedex 05, France
[2] CNRS UMR 7614, Laboratoire de Chimie Physique-Matière et Rayonnement, 11 rue Pierre et Marie Curie, F-75231 Paris cedex 05, France
[3] CNRS - ICMMO, Université Paris Sud, bât 410, F-91405 Orsay Cedex
[4] CMTC Grenoble INP, Bât PHELMA Campus, BP75 - Domaine Universitaire, F-38402 Saint Martin d'Hères, France
[5] BRGM, 3 Ave C. Guillemin, BP 36009, F-45060 Orléans cedex 2, France
[6] Microscopie Icaunaise, 6 Berthellerie, F-89770 Bœurs en Othe, France

We conducted an inter-laboratory study of a metallic glass whose main component is nickel. Two determinations of the mass fractions of the different elements present within the sample were asked to the participants: one at an acceleration voltage of 15 or 20 kV and another one at 5 kV. We compare the mass fractions obtained from wavelength dispersive (WDS) and energy dispersive spectrometries (EDS) and also try to find an influence of the kind of EDS detector and its entrance window, the background subtraction method, the use or not of standards as well as the quantification method. Both means of WDS and EDS mass fractions are close to the reference values. The dispersion of the results was larger at 5 kV than at 15-20 kV owing to the use of the L lines rather than K lines and to the lowest collected intensities. There is an exception with the case of boron because at the lowest voltage, the excitation condition is more favourable for the production of the K line. It appears that the dispersion of the results is larger with EDS than with WDS but it was not possible to find a correlation between the large dispersion and one of the considered experimental parameters and quantification factors. Thus, one can think that electron microprobes are inherently better for the determination of mass fractions or that the implementation of quantitative analysis must be optimized for some cases, especially in SEMs.

# Introduction

Quantitative x-ray microanalysis can be performed either on scanning electron microscopes (SEM) or electron microprobes (EPMA), either with wavelength dispersive spectrometry (WDS) or energy dispersive spectrometry (EDS). Then, depending on the experimental constraints, quantitative analysis can be done either with high (a few 10 keV) or low (a few keV) energy electrons, either with standard materials or without standard. Thus, it is difficult to find some laboratories working in the same conditions. Consequently, the practitioners of quantitative x-ray microanalysis cannot know the reliability of their measurements and quantifications owing to the lack of comparison. Thus, they should rely on samples whose composition is certified by a metrology institute or other quantification methods.

This is why, mainly under the auspices of GN-MEBA [1], we conducted this inter-laboratory study of a metallic glass, in order for the users to know if they introduce any systematic error in their results or if they use the best-suited experimental conditions and data treatments. The results of the participating laboratories were kept anonymous. This study also points to some inherent difficulty regarding the inaccuracy of some fundamental parameters [2]. Let us note that a recent inter-laboratory study of steel using WDS and EDS at low voltage has given evidence of inaccuracy of mass absorption coefficients for the L lines of the *3d* elements [3]. Let us mention other inter-laboratory comparison studies based on WDS [4] or EDS [5] and involving less than 10 participants. We would like to emphasize that the authors of the paper are not members of a certification organism and that the aim of this inter-laboratory study is not to evaluate or give a score to the participating laboratories. We only intent to give to the participating laboratories the means to compare their measured values obtained on a well-characterized sample with those of other laboratories working a priori in the same or very similar experimental conditions. So, we do not follow the protocol of ISO [6] or IUPAC [7] regarding the proficiency testings.

# Sample

The pristine sample is a thin tape of a metallic glass with Ni as the major component, whose size is 15 cm x 2 cm. Its thickness is 0.3 mm, *i.e.* much larger than the penetration length of the most energetic electrons used in this study. It was prepared by melt spinning to avoid its crystallization and thus was amorphous. The amorphous state of the sample was confirmed by electron backscattered diffraction. The mass fraction of the different elements, as determined by ICP-MS, was: B, 3.6%; Cr 10.4%; Fe 5.6%; Co, 23.4%; Ni, 50.3%; Mo, 6.7%. Al and Si were found as impurities in the 0.1-0.2% mass fraction range. These values are given with a 4% measurement uncertainty. Owing to the homogeneous nature of the sample, the ICP-MS determinations, requiring the dissolution of the sample, are valid as references for the x-ray analyses. SEM images have shown that the sample is homogeneous at the micrometer scale. A WDS mapping of the K$\alpha$ emission of B, Cr, Fe, Co and Ni as well as of the L$\alpha$ emission of Mo has demonstrated the uniform distribution of these elements. Pieces of 5 mm x 5 mm were supplied to laboratories interested in participating to this inter-comparison study. Being conductor, no dedicated preparation was required except the cleaning of the surface.

## Laboratories and operating conditions

The number of participating laboratories is 56, involving a total of 69 electron columns and 75 spectrometers. The details are the following:
- 69 electron columns:  - 56 SEMs divided into 28 FEG, 27 W and 1 LaB$_6$;
  - 13 electron microprobes;
- 75 spectrometers:  - 57 EDS divided into 37 SDD and 20 Si(Li);
  - 18 WDS installed on 13 electron probes and 5 SEMs.

The meaning of the different terms is: FEG, field emission gun; W, tungsten filament gun; LaB$_6$, lanthanum boride filament gun; SDD, silicon drift detector; Si(Li), lithium drifted silicon diode. We note, with respect to previous inter-laboratory studies initiated by GN-MEBA in 2005 and 2009, that no Ge detector was used in this study whereas the proportion of SDD is increasing at the detriment of Si(Li) detectors. The number of electron microprobes is stable while the number of WDS apparatus on SEM increases.

The only instruction given to participants was to make an analysis at high voltage, 15 or 20 kV, and another one at 5 kV. The other experimental conditions were left free but it was asked to specify them. If the user did not know one particular condition, it is noted as undefined. For every different condition, each measurement was repeated at least ten times in a row, changing the location on the sample each time and without new calibration between successive measurements. The laboratories returned the mean value and the standard deviation.

# Results

In the following, we use the arithmetic mean and the square root of the variance to present the results. They will be given with one standard deviation, *i.e.* with a 68% probability to find the consensus value derived from participants in the confidence interval. All the results were accepted provided that the participants gave the details of their experimental, treatment and quantifying details as listed below.

## EDS analysis at 15 or 20 kV

In this case the quantification is based on the analysis of the K lines of the *3d* elements and boron and L lines of molybdenum. The expected difficulties regarding this analysis are some interferences: Fe K$\beta$ with Co K$\alpha$ and Co K$\beta$ with Ni K$\alpha$. Let us also mention that the Mo L$\alpha$ line could be misinterpreted as the S K$\alpha$ line. The detection and analysis of B K$\alpha$ emission (185 eV [8]) could also be problematic owing to the very high overvoltage (the ratio of the electron energy to the ionization threshold energy), in the 75 – 100 range, and to an interference with the intense Mo M$\zeta$ line (193 eV [8]). Another problem could originate from the determination of the continuous background, which is generally worse toward the lowest photon energies (<1 keV) owing to the poor transmission of these x-rays through the detector window. With the 57 spectrometers, 64 analyses were performed with the following details:

- analysis mode: 36 scanning mode, 19 spot mode, 1 defocused electron beam analysis and 8 undefined; this will not be discussed in the following owing to the sample homogeneity;
- detector: 22 Si(Li) and 42 SDD;
- entrance window: 59 polymer, 3 Be and 2 undefined (but most probably polymer);
- background determination: 31 model, 20 numerical filter and 13 undefined;
- standard: 25 from library, 23 without, 14 real and 2 undefined;
- quantification method: 22 ZAF method [9–11], 50 $\varphi(\rho z)$ approach [9,11–14] and 12 undefined.

In these conditions, Cr, Fe, Co, Ni and Mo have been detected and quantified by all the laboratories, whereas only 38 treated Si, 28 Al and 22 B. We present in Table 1 the weight concentrations calculated as the mean of all the measurements.

Table 1: Mass fraction of the different elements present within the metallic glass sample analysed at 15 or 20 kV with the EDS apparatus. Reference values are compared to the mean of the measured values. The range of the experimental determinations is also given.

| Element | B | Cr | Fe | Co | Ni | Mo |
|---|---|---|---|---|---|---|
| Reference (wt%) | 3.6 | 10.4 | 5.6 | 23.4 | 50.3 | 6.7 |
| Experimental mean (wt%) | 4.4 ± 3.0 | 10.5 ± 0.6 | 5.4 ± 0.4 | 23.5 ± 0.8 | 51.4 ± 2.1 | 7.1 ± 1.4 |
| Experimental range (wt%) | 1.2 – 10.6 | 9.1 – 12.5 | 2.9 – 5.9 | 21.3 – 25.3 | 45.4 – 56.9 | 4.2 – 14.1 |

It appears that the mean of the measured value is quite close to the reference value, even for the boron concentration whose determination was anticipated as challenging. However, the dispersion of the results is quite large, boron showing the widest range of concentration. Within the statistical uncertainty, it has not been possible to observe an influence of the nature of the detector, SDD or Si(Li), the choice of the standard, the method to subtract the continuous background or the quantification method.

### EDS analysis at 5 kV

In this case, the quantification is based on the analysis of the L lines except for boron analysed with the K line. The expected difficulties arise owing to the strong interferences between the Cr, Fe, Co and Ni lines, the low intensity of the Mo L lines coming from the small overvoltage, 2, and again the B K$\alpha$ – Mo M$\zeta$ interference. The detection and analysis of the B K$\alpha$ emission should be less problematic because its overvoltage drops to 25. As an example, we show an EDS spectrum in Figure 1.

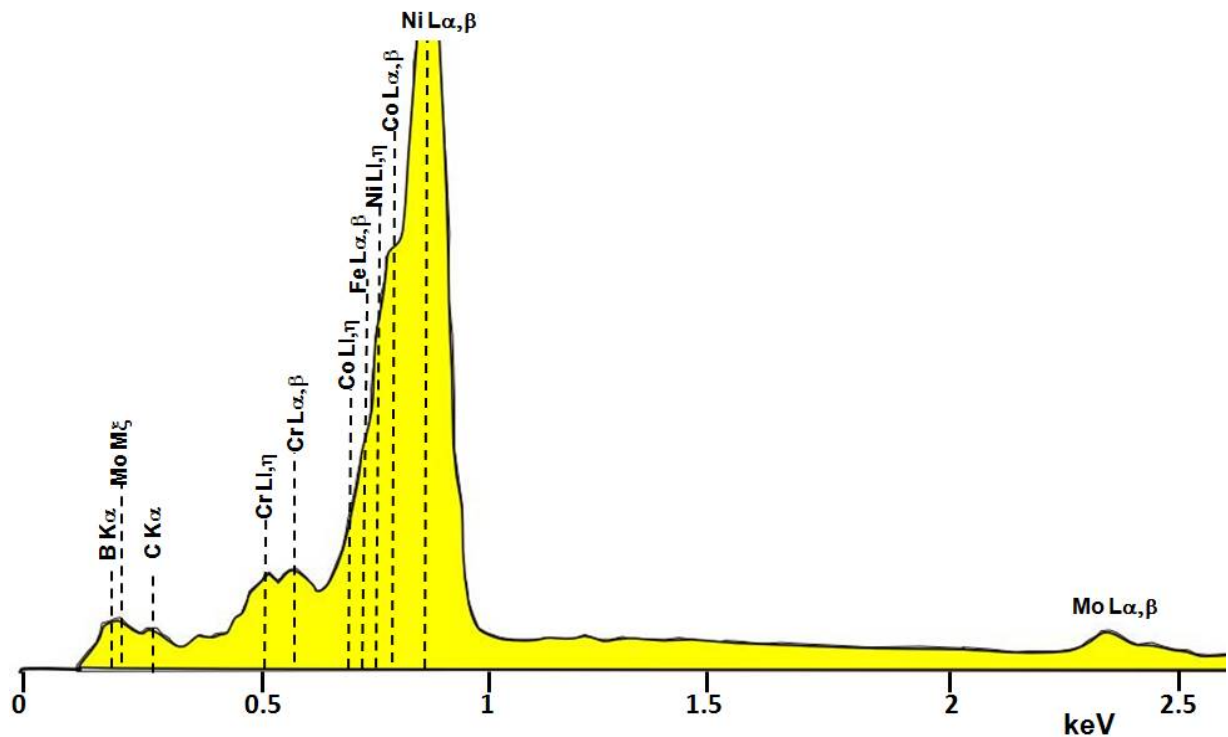

Figure 1: EDS spectrum of the metallic glass sample obtained at 5 kV.

In this case, only 43 analyses have been performed. All have detected Ni and Mo, but only 42 Co, 41 Cr, 37 Fe and 21 B. This is most probably due to some deconvolution misinterpretation, but could also come from a poor quality spectra and then a faulty automatic identification procedure. The determined concentrations are given in Table 2. Once again, there is a fair agreement between the mean of the measured values and the reference ones. However, with respect to the analysis at 15-20 kV, the accuracy of the results is lower and their dispersion larger. There is an exception for the case of boron owing to the improved ionization of B K level with a 25 overvoltage. For Ni and Mo concentrations, two aberrant measurements have been taken out of the dataset. One is particularly surprising as it gives calcium as the major element whereas no Ca peak was observed on the EDS spectrum. Most probably this was caused by a lack of control of the background subtraction by the laboratory and to a poor spectrum quality due to low count, although it should be easy to treat such a case as the background is very flat in this energy range and with this sample. This points out the interest of really studying any spectra and not only push on the "quant" button and let the software operate automatically.

Table 2: Mass fraction of the different elements present within the metallic glass sample analysed at 5 kV with the EDS apparatus. Reference values are compared to the mean of the measured values. The range of the experimental determinations is also given. The lines "Model" and "Filter" give the concentrations determined by using a model or a filter to remove the background.

| Element | B | Cr | Fe | Co | Ni | Mo |
|---|---|---|---|---|---|---|
| Reference (wt%) | 3.6 | 10.4 | 5.6 | 23.4 | 50.3 | 6.7 |
| Experimental mean (wt%) | 3.1 ± 2.0 | 10.5 ± 3.8 | 3.9 ± 2.8 | 22.8 ± 5.3 | 54.0 ± 4.7 | 6.0 ± 1.9 |
| Experimental range (wt%) | 0.5 – 7.7 | 0 – 20.8 | 0 – 9.4 | 0 – 31.7 | 45.3 – 64.3 | 3.3 – 14.6 |
| Model (wt%) | 3.4 ± 2.4 | 11.4 ± 3.6 | 3.5 ± 3.2 | 21.7 ± 3.9 | 55.1 ± 4,9 | 6.2 ± 2.5 |
| Filter (wt%) | 3.5 ± 1.7 | 10.6 ± 2.4 | 4.9 ± 1.9 | 25.2 ± 2.3 | 52.9 ± 3.4 | 5.9 ± 1.5 |

The different considered parameters do not have a significant influence on the results, except for the method used to remove the background. When the numerical filter is used, the accuracy of the measurements is better and their dispersion is lower, see Table 2. It should also be noted that among the 6 laboratories, which were not able to quantify iron, 5 were modelling the background (the 6[th] did not mention which subtraction method was used).

However, this looks more like a faulty deconvolution procedure owing to the low Fe concentration and to the presence of nearby Co and Ni lines (see Figure 1).

**WDS analysis at 15 or 20 kV**

On electron microprobes, the analysis of the different emission lines was done by using a Ni/C (12 times) or a Mo/B4C (once) multilayer for B K$\alpha$, PET (pentaerythritol, 7 times) or LiF (lithium fluoride, 5 times) crystal for Cr K$\alpha$, a LiF crystal for Fe, Co and Ni K$\alpha$, and a PET crystal for the Mo L$\alpha$. One user gave undefined conditions. The participants detected all the elements.

The operating conditions for the WDS spectrometers on SEM are presented in Table 3. It can be seen that only 2 laboratories were able to detect boron. Sometimes, K$\alpha$ or L$\alpha$ lines were used by the different laboratories to measure the same element. We could have expected the use of a coupling of WDS and EDS measurements where EDS analysed the major elements and WDS the minor ones, but this was not done by the participants in this inter-laboratory study. A specific study involving coupling of EDS and WDS spectrometries should be proposed to test the capabilities of such an approach.

Table 3: Operating conditions for WDS/SEM apparatus for the analysis at 15 or 20 kV: choice of the crystal or multilayer noted "Multi", and of the analysed line. TAP is the thallium acid phthalate crystal.

|       | B        | Cr        | Fe        | Co        | Ni        | Mo        |
|-------|----------|-----------|-----------|-----------|-----------|-----------|
| Labo1 | -        | TAP L$\alpha$ | TAP L$\alpha$ | TAP L$\alpha$ | TAP L$\alpha$ | TAP L$\alpha$ |
| Labo2 | -        | TAP L$\alpha$ | TAP L$\alpha$ | TAP L$\alpha$ | TAP L$\alpha$ | EDS       |
|       | -        | EDS       | TAP L$\alpha$ | TAP L$\alpha$ | TAP L$\alpha$ | EDS       |
|       | -        | EDS       | EDS       | TAP L$\alpha$ | TAP L$\alpha$ | EDS       |
| Labo3 | -        | LiF K$\alpha$ | LiF K$\alpha$ | LiF K$\alpha$ | LiF K$\alpha$ | PET L$\alpha$ |
| Labo4 | Multi K$\alpha$ | LiF K$\alpha$ | LiF K$\alpha$ | LiF K$\alpha$ | TAP L$\alpha$ | PET L$\alpha$ |
| Labo4 | Multi K$\alpha$ | LiF K$\alpha$ | LiF K$\alpha$ | LiF K$\alpha$ | LiF K$\alpha$ | PET L$\alpha$ |

Table 4 shows the mass fraction deduced from the measurements on the electron microprobes and WDS spectrometers on SEMs, in comparison to the reference concentrations. The procedures used for matrix corrections were the ZAF method [9–11] or the φ(ρz) approach [9,11–14]. Both kinds of measurements give concentrations close to the reference ones. However, electron microprobe is more precise and also leads to less dispersed results. This result could be related to several EPMA technical designs and specificities, especially beam current stability and accuracy of the sample Z-positioning with respect to the geometry of WDS spectrometers. As a consequence, the composition of the sample is determined to a few tenths of percent. With respect to the EDS determination, more precise results are obtained by WDS regarding the determination of the boron concentration, the concentration of the other elements being correctly evaluated by both WDS and EDS techniques. However, with respect to the dispersion of the results, EDS gives the worse results, as is illustrated in Figure 2 where the EDS and electron microprobe quantifications of Ni and Mo are compared. These results clearly illustrate the interest of WDS compared to EDS for the accuracy of quantitative x-ray measurements in electron microscopy and for the quantification of light elements.

Table 4: Mass fraction of the different elements present with the metallic glass sample analysed at 15 or 20 kV with the electron microprobe and WDS spectrometer on SEM. Reference values are compared to the mean of the measured values. The range of the experimental determinations is also given.

|  | Element | B | Cr | Fe | Co | Ni | Mo |
|---|---|---|---|---|---|---|---|
|  | Reference (wt%) | 3.6 | 10.4 | 5.6 | 23.4 | 50.3 | 6.7 |
| Electron probe | Exp. mean (wt%) | 3.8 ± 0.7 | 10.4 ± 0.2 | 5.4 ± 0.1 | 23.3 ± 0.3 | 50.6 ± 0.8 | 6.7 ± 0.3 |
| | Exp. range (wt%) | 2.5 – 5.1 | 10.1 – 10.6 | 5.2 – 5.7 | 22.5 – 23.6 | 49.5 – 52.2 | 6.0 – 7.1 |
| WDS/ SEM | Exp. mean (wt%) | 3.4 ± 0.9 | 10.1 ± 1.7 | 5.2 ± 0.7 | 23.3 ± 0.3 | 54.7 ± 4.3 | 6.3 ± 1.0 |
| | Exp. range (wt%) | 2.8 – 4.0 | 7.3 – 12.0 | 4.0 – 5.9 | 19.7 – 25.6 | 50.6 – 61.7 | 4.7 – 7.0 |

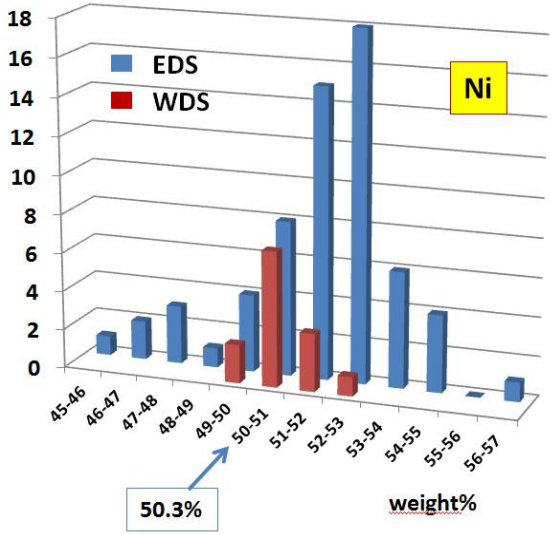
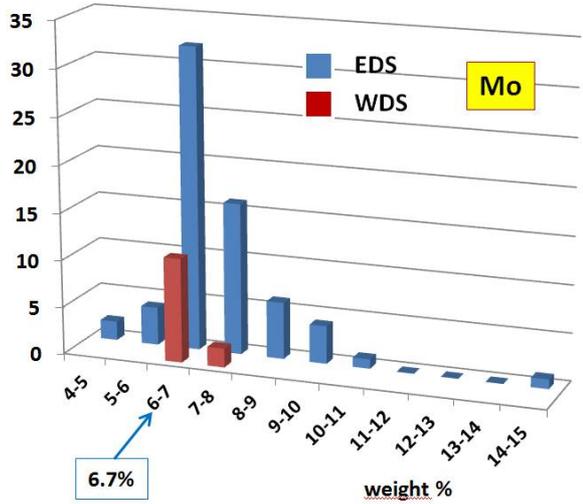

Figure 2: For Ni and Mo, dispersion of the mass fractions deduced from EDS and WDS measurements. The values in rectangular boxes correspond to the reference mass fractions.

**WDS analysis at 5 kV**

With this operating condition, only 7 electron microprobe and 3 WDS/SEM analyses were done. Among the 3 last ones, only one quantified boron. As in EDS, the L$\alpha$ line of the transition metals were used for quantification, the acceleration voltage being too low to allow exciting K shells of *3d* and *4d* elements. As can be seen if Figure 3, owing to the improved spectral resolution in WDS analysis, the intensity maximum of the B K$\alpha$ and Mo M$\zeta$ emissions are well separated. This makes easier the boron quantification. The mass fractions of the different elements are given in Table 5. Owing to the small number of analyses, the dispersion of the results is not given. As was observed in the EDS case, the standard deviation on the mass fraction increases when the acceleration voltage decreases, except for boron.

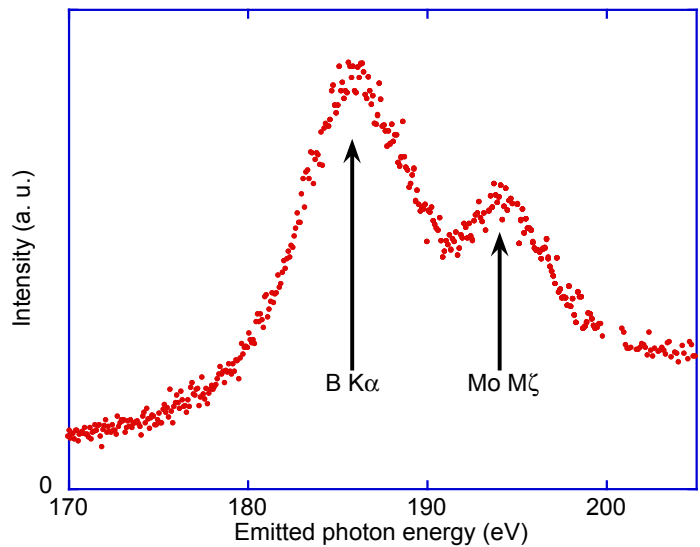

Figure 3: WDS spectrum of the metallic glass sample in the range of the B K$\alpha$ emission. It is obtained with an electron microprobe working at 5 kV and working with a Ni/C multilayer.

Table 5: Mass fraction of the different elements present within the metallic glass sample analysed at 5 kV with the electron microprobe and WDS spectrometer on SEM. Reference values are compared to the mean of the measured values. Boron value obtained by WDS on SEM is given without standard deviation because only one determination was done.

|  | Element | B | Cr | Fe | Co | Ni | Mo |
|---|---|---|---|---|---|---|---|
|  | Reference (wt%) | 3.6 | 10.4 | 5.6 | 23.4 | 50.3 | 6.7 |
| Electron micro-probe | Exp. mean (wt%) | 3.7 ± 0.5 | 10.0 ± 1.3 | 4.8 ± 1.3 | 26.0 ± 2.6 | 56.8 ± 6.1 | 7.1 ± 0.9 |
| WDS/ SEM | Exp. mean (wt%) | 2.5 | 7.7 ± 1.8 | 5.3 ± 2.2 | 26.3 ± 3.3 | 60.7 ± 3.4 | 6.8 ± 0.3 |

It appears at 5 kV with WDS on SEM and with a microprobe that the boron and molybdenum concentrations are well evaluated, the one of iron is under-evaluated but within the uncertainty whereas the cobalt and nickel ones are over-estimated. The determination for chromium is correct with the electron microprobe and under-estimated with the WDS/SEM. The origin of the over-estimation of the Co and Ni concentration can be due to some problem in the correction factors, particularly owing to a large uncertainty in the mass absorption coefficients when using the L lines, as recently emphasized in Ref. [3]. Indeed from the measured intensity on the sample, a k-ratio is determined by dividing it by the intensity of the standard. Then, the mass fraction is deduced from the k-ratio thanks to a mathematical model taking into account matrix, absorption and fluorescence effects, which are not the same in the studied sample and in the reference. The problem of uncertainty in mass absorption coefficients does not exist for the 20 kV analysis with K lines.

## Discussion and conclusion

From this inter-laboratory study, it appears that the dispersion of the mass fractions is rather large. This shows the utility of such a study as we demonstrate that the implementation of quantitative analysis must be optimized for some cases. For example, in the WDS/SEM analysis, there was no combined analysis where EDS could be used to determine the concentration of the major elements. However, the value resulting from the mean of all the determinations is always in good agreement with the reference value.

We found that the electron microprobes give better results than EDS spectrometer, both from the point of views of accuracy (see Table 6) and dispersion of measurements. Owing to small number of WDS systems on SEMs, it is difficult to obtain a conclusive comparison with electron microprobes and EDS systems. None of the envisaged experimental and quantification factors can be regarded as the responsible of the better performances of electron microprobes over EDS. To illustrate this point, we present in the Table 6 a comparison of the means of the uncertainties of the EDS and WDS quantifications made at 20 kV. Each laboratory calculates its standard deviation from the 10 measurements made for each determination; then, the mean is calculated by taking into account all the laboratory values. Then, we calculate the ratio of this mean to the reference mass fraction to get what we call the relative standard deviation. Proceeding this way, it can be seen that WDS technique is twice more precise than the EDS one. Thus, one can think that electron probes are inherently better for the determination of mass fractions. These types of analysis show also that EDS analysis is not as straightforward as sometimes expected and that users need good training to fully exploit their equipment and results. As seen, the best results are coming from WDS. This can be explained by several reasons including:

- a WDS spectra is generally easier to treat. Indeed, its shape generates a better resolution, a better signal/noise ratio, a more flat background, so all these facilitate identification and quantification.
- as WDS is a technique that requires standards analysis, it could be possible that WDS users are more aware of the quantification process than some EDS users and then can give more accurate or less disperse results.

However, we can expect that following some recommendations of Newbury and Ritchie [15], such as examine every analysis, understand the particular procedures of the standardless analysis, try to know how the commercial program provide their results, …, then the quality of the EDS results could approach the one obtained through WDS.

Table 6: For the EDS and WDS determinations of the mass fractions, mean values of the uncertainties and calculation of the ratio of this mean to the reference value.

| Element | Reference mass fraction (wt%) | Mean value of the EDS standard deviation | EDS relative standard deviation (%) | Mean value of the WDS standard deviation | WDS relative standard deviation (%) |
|---|---|---|---|---|---|
| Cr | 10.4 | 0.11 | 1.06 | 0.05 | 0.48 |
| Fe | 5.6 | 0.08 | 1.43 | 0.04 | 0.71 |
| Co | 23.4 | 0.18 | 0.77 | 0.11 | 0.47 |
| Ni | 50.3 | 0.2 | 0.40 | 0.23 | 0.46 |
| Mo | 6.7 | 0.13 | 1.94 | 0.07 | 1.04 |

Although most of the WDS analysis gave results close to the reference values of the chemical composition, there was a problem with the analysis at 5 kV, which leads to an over-estimation of the Co and Ni mass fractions. This raises the question of the accuracy of the fundamental parameters, particularly of the mass absorption coefficient of *3d* elements when the L lines are used [3], but also how the quantification programs are working. For example regarding this last point, in the EDS analysis at 15 or 20 kV, it appears that some quantification programs return too high values of the k-ratios but that the deduced mass fractions are correct. This suggests that these programs do not use the "conventional" definition of the k-ratio. Then, comparison between k-ratio determinations performed on different systems can be more or less illusory. Thus, we point out the necessity to get information from EDS and WDS suppliers regarding the data treatment procedures in their quantification programs.

The full analysis of the results of this inter-laboratory test can be found (in French) at the following address: http://www.gn-meba.org/ech_tests.htm.

## Acknowledgments
Members of the GN-MEBA, SEMPA users group and Belgian laboratories coordinated by Université Catholique de Louvain are thanked for their participation to this inter-laboratory study.